%% file: asm07.tex
\documentclass[footinfo,envcountresetchap]{svmult}
\usepackage{graphicx}         % to include .eps figures
\usepackage{color}
\usepackage[bottom]{footmisc} % for proper footnotes

\usepackage{epsfig}
\usepackage{psfrag}
\usepackage{amsfonts}          % |
\usepackage{amssymb}           % | AmS-LaTeX packages
\usepackage{amsmath}           % |
\usepackage{ulem}

\newcommand{\f}[2]{\frac{#1}{#2}}
\newcommand{\dd}{\partial}
\newcommand{\de}{{\rm \, d}}

\renewcommand{\Vec}[1]{\mbox{\boldmath $ #1$}}
\renewcommand{\v}{\Vec}

\allowdisplaybreaks

\input{abbr.tex}

\newcommand{\subsub}[1]{\noindent {\bf #1.}}

\newcommand{\reva}[2][0]{#2}
\newcommand{\vnba}[1]{#1}

\newcommand{\vnbb}[1]{#1}

\newcommand{\sizedel}[1]{}
\newcommand{\sizerepl}[2]{#2}

\newcommand{\vnbdel}[1]{}
\newcommand{\vnbrepl}[2]{#2}

\newcommand{\new}[1]{#1}


\renewcommand{\emph}{\textit}
\renewcommand{\em}{\it }

\begin{document}

\title{Analytically Solvable Asymptotic Model of Atrial Excitability}

% Additionally, use \titlerunning for an abbreviated version of
% your title (only if the full title is too long to be displayed 
% in the header line of the following pages), else comment out
\titlerunning{Model of Atrial Excitability}

\author{R.D.~Simitev \and V.N.~Biktashev}

% Additionally, use \authorrunning if you have more than 2 authors, 
% else comment out
\authorrunning{Simitev \& Biktashev}

\institute{
Applied Mathematics, Department of Mathematical Sciences, 
 The University of Liverpool, Liverpool L69 7ZL, UK,
\texttt{Radostin.Simitev@liverpool.ac.uk}}

\maketitle

\begin{abstract}
We report a three-variable simplified model of excitation fronts in human
atrial tissue. The model is derived by novel asymptotic techniques
\new{from the biophysically realistic model of Courtemanche \etal{}
\cite{CRN98}} in extension of our previous similar models.
An iterative analytical solution of the model is presented 
\new{which is in excellent quantitative agreement with the realistic
model.}
It opens new possibilities for analytical studies as well as for
efficient numerical simulation of this  and other cardiac models of
similar structure.  
% please provide a list of keywords for your work
\index{cardiac modelling} \index{asymptotics} \index{excitation}
\index{wave front}\\[3mm]
Published \textbf{in "Mathematical Modeling of Biological Systems" ,
  Vol 2, A. Deutsch, R. Bravo de la Parra, R. de Boer et al. (eds.)
  Birkhaeuser, Boston, pp. 289-302, DOI:10.1007/978-0-8176-4556-4\_26,
  2008.}
\end{abstract}

%%%%%%%%%%%%%%%%%%%%%%%%%%%%%%%%%%%%%%%%%%%%%%%%%%%%%%%%%%%%%%%%%%%

%%%%%%%%%%%%%%%%%%%%%%%%%%%%%%%%%%%%%%%%%%%%%%%%%%%%%%%%%%%%%
%     PAPER TEXT                                            %
%%%%%%%%%%%%%%%%%%%%%%%%%%%%%%%%%%%%%%%%%%%%%%%%%%%%%%%%%%%%%

\section{Introduction}
\label{sec:1}

The mechanical activity of the heart is controlled by electrical
impulses propagating regularly \vnbrepl{\new{on its surface}}{through it} during our entire
lifespans \cite{Mohrman}. A disturbance in the regular propagation
may lead to life-threatening \emph{cardiac arrhythmias} \index{cardiac arrhythmias}
\cite{Spooner}. Sudden cardiac death, for instance, accounts for 300
000 to 400 000 deaths annually in the United States alone
\cite{EngelsteinZipes,Myerburg}, \ie{} more than AIDS, breast and lung
cancer.
This entails intensive research into the mechanisms of heart
functioning and failure. The accumulated information
reveals an overwhelming complexity of the patterns of
electrical cardiac activity. 

True understanding of the experimental
data requires the development of appropriate qualitative cardiac models
\cite{Clayton,Holden-Panfilov-1997,Kohl-etal-2000}.  
One approach to cardiac modelling is to
take into account the various levels of membrane, cellular and
myocardial structure and their interactions and to model the
\emph{action potential} \index{action potential} (AP) on the basis of 
% \reva[R1]{experimental measurements of}
\new{experimental measurements of}
ion fluxes in as much detail as possible. \new{The resulting} models
are known as \emph{realistic} \vnbdel{cardiac models} or
\emph{detailed ionic models}\index{realistic cardiac models}.
%- \reva[R1]{They are capable of reproducing the full body of
%- experimental data available to date for the specific cardiac tissue in
%- question and thus we regard them as fundamental governing
%- equations. For instance, in our study of the excitability of human
%- atrial tissue we start from the well-established in the literature
%- realistic model of Courtemanche \etal{} \cite{CRN98}.} 
The first example of this type of models was developed by Noble
\cite{Noble1960,Noble-1962} and there are now such models for various
cardiac cells in different species,   \eg{}
\cite{BeelerReuter,Demir,LuoRudy,Nygren,Varghese} and many 
others. However, since these models are very complex and highly 
nonlinear, it is difficult to 
\vnbrepl{evaluate the}{assess} contribution of specific
model components to different patterns of activity. Furthermore, their
computation is arduous because they contain a large
number of equations and small parameters. They become very
expensive and time-consuming especially when large volumes of tissue
are simulated. 
One possible alternative is to search for simplified
\vnbrepl{dynamical systems}{models}\index{simplified models}
which could mimic the most important AP properties, allow analytical
studies and reduce the computing requirements. 
Many simplified models have been suggested, either phenomenologically,
or based on the structure of the realistic models,
\cite{AlievPanfilov1996,%
      Bernusetal2002,%
      DuckettBarkley2000,%
      FentonKarma1998,%
      Hinch2002}
\etc{}. 
However, all of these models contain arbitrary elements in
the sense that they are not derived from any of the realistic
biophysical cell models and \vnbdel{so they} lack explicit correspondence with
the biophysical structure 
of the cardiac tissues. 
\vnbrepl{
\eg{} van der Pol and van der Mark modelled
heartbeat \cite{vanderPol-vanderMark-1928} as an relaxation
oscillator.
}{  
E.g. van der Pol and van der Mark modelled
heartbeat as an \vnbb{electronic} relaxation
oscillator\cite{vanderPol-vanderMark-1928}.  
}

\begin{figure}[t]
\begin{center}
%/export/simitev/PAPERS.PLOTS/04_Dresden/REVISION 
\psfrag{eqref}{\hspace{-3mm}\footnotesize Eq.~\eqref{e:0020}}
\psfrag{x}{\footnotesize $x$}
\psfrag{t=0}{\scriptsize $t$=0}
\psfrag{t=8}{\scriptsize $t$=8}
\psfrag{t=16}{\scriptsize $t$=16}
\psfrag{t=24}{\scriptsize $t$=24}
\psfrag{t=32}{\scriptsize $t$=32}
\psfrag{t=38}{\scriptsize $t$=38}
\psfrag{t=40}{\scriptsize $t$=40}
\psfrag{t=76}{\scriptsize $t$=76}
\psfrag{t=80}{\scriptsize $t$=80}
\psfrag{t=114}{\scriptsize $t$=114}
\psfrag{t=120}{\scriptsize $t$=120}
\psfrag{t=160}{\scriptsize $t$=160}
\psfrag{t=152}{\scriptsize $t$=152}
\epsfig{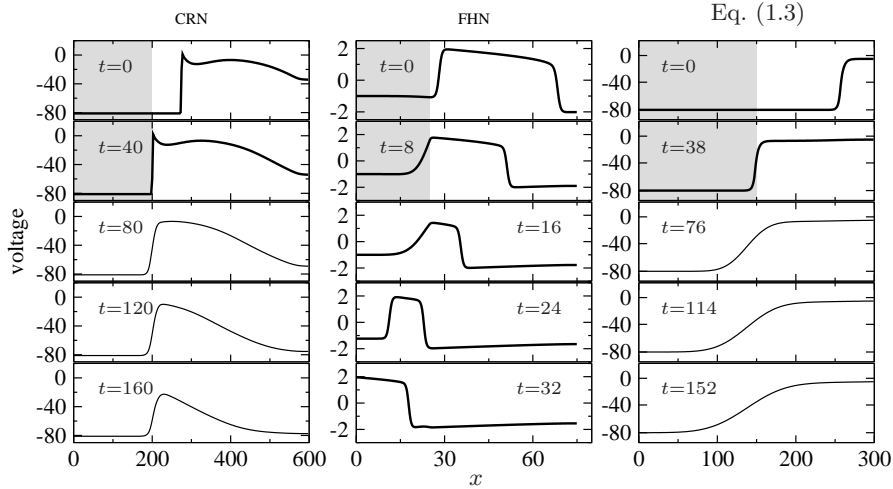} 
\end{center}
\caption[]{\reva[R2]{Propagation of excitation in the models 
  of Courtemanche \etal{} \cite{CRN98} (first column), FitzHugh-Nagumo (second column)
  and in equations \eqref{e:0020} (third column), through a temporary block
  of excitability, introduced by artificially reducing the value of
  a parameter representing the main excitatory ionic current responsible for the
  initiation of the front. E.g.~in the three-variable model
  \eqref{e:0020} this parameter is $j$ which was decreased from the
  normal value of 0.9775 to 0.28 during the block.   In FHN,
  propagation resumes after the block is removed; in CRN and
  \eqref{e:0020} it does not. 
}
}
\label{f:0000}
\end{figure}

Our recent studies
\cite{Biktasheva-etal-2003,%
Suckley-Biktashev-2003a,%
Suckley-Biktashev-2003b,%
Biktashev-Suckley-2004}  have demonstrated a serious disadvantage of
the most popular and successful simplified generic model of cardiac excitability
--  the FitzHugh-Nagumo (FHN) equations \cite{FitzHugh-1961,Nagumo-etal-1962}.
It cannot describe adequately the one feature of excitation
propagation which is most important for medical applications, namely
the way regular propagation fails and arrhythmias occur.  An excitation
wave in real atrial tissue, or in a realistic model, may fail to
propagate if the temporal gradient of the transmembrane voltage at the
front becomes too small to excite the tissue ahead of it \eg{} if the
wave fails to propagate fast enough \cite{Biktasheva-etal-2003}. Then
the wave front looses its sharp spatial gradient and its further
spread is purely diffusive, \ie{} the \vnbb{front} dissipates. This happens
long before the back of the excitation impulse catches up with the
front. 
\vnbdel{\reva[R2]{This mechanism of front dissipation is illustrated in
the \vnba{first} column of figure \ref{f:0000} for the Courtemanche \etal{}
(CRN) \cite{CRN98} realistic model of atrial tissue.}} 
This type of propagation failure does not exist in a FitzHugh-Nagumo type model
\cite{Biktashev-2003} since it is known that the propagation of a wave
front in this system may be slowed down, halted or even reversed
\cite{Fife-1976}. 
%\vnbdel{\reva[R2]{Indeed, the \vnba{second} column of figure
%  \ref{f:0000} shows that when a temporary block of excitability is
%  imposed the front of a FitzHugh-Nagumo wave is halted but does not
%  dissipate. After the block is removed the wave continues its propagation.
%}}
\reva[R2]{
  This phenomenon is illustrated in figure~\ref{f:0000}: a temporary
  block of excitability only temporary halts the FHN wave, but
  completely disrupts propagation in the realistic model, even though
  it lasts much shorter than the AP.}

\sizedel{Earlier we have proposed a simplified model of wave front propagation
in atrial tissue derived from the \sizerepl{model of Courtemanche \etal{}}{CRN model}
\cite{CRN98} by novel asymptotic methods \cite{Biktasheva-etal-2005}.
\reva[R2]{This simplified model was capable of reproducing the front dissipation behaviour of the
realistic model.}  It was based on the observation
\cite{Biktashev-2002,Biktashev-2003} that propagation of excitation and its block 
can be successfully predicted by a simplified model of the excitation
front, obtained by an asymptotic description focused on the fast
sodium current, $\INabar$. That was a two-variable model taking into
account only the inactivation gate $h$ of the sodium current. It admitted
an exact analytical solution for a piecewise-linear `caricature' 
which for an appropriate choice of parameters reproduced the key
qualitative features of  the system. }

\sizerepl%
{We have recently found that for good quantitative
predictions, one must also take into account not only the dynamics of
$\E$ and $h$, as in the in the simplified model of
\cite{Biktashev-2002,Biktashev-2003}, but also the influence of the
$m$ activation gate \cite{Simitev06b}.} 
{Earlier we have proposed novel asymptotic methods of reduction of
cardiac equations \cite{Biktasheva-etal-2005}. }  
In this paper we use these methods to derive a
three-component description of the propagating excitation fronts and
their dissipation. \reva[R2]{The  virtue of our model is that it reproduces
propagation failure unlike the FitzHugh-Nagumo models because it is
derived in a reliable way from the  realistic ionic model.}
We also report an analytical solution for this three-component model. The
analytical solution is constructed as an iterative procedure and it
may be seen as a generalisation in which the caricature solutions
\sizerepl{}{presented in our earlier papers}
\cite{Biktashev-2002,Biktashev-2003,Simitev06b} appear as first  approximations.  
\reva[R4]{This, is to our knowledge, the first analytical solution,
albeit in quadratures, which gives a numerically accurate prediction
of the front propagation velocity (within 16\%) and its profile
(within 0.7 mV) in human atrial tissue. \vnbdel{Furthermore the evaluation of
the analytical solution is far more effective than a direct solution
of the 21-equation PDE system of the realistic model \cite{CRN98}.}}

\section{Mathematical formulation of the problem} 
\label{sec:2}

\subsub{Atrial tissue model}
In our study atrial tissue is a one-dimensional, homogeneous and
isotropic medium satisfying a system of \emph{reaction--diffusion
equations} \index{reaction--diffusion equations}
\begin{gather}
\D_{T} \u = \hat{\v D} \cdot \D_{X}^2 \u + \v F(\u)
\label{n:0010}
\end{gather}
where  $\v F(\u)$ \vnbb{is} a vector defined according to the atrial single-cell
realistic \sizerepl{model of Courtemanche \etal{}}{CRN model} \cite{CRN98}, 
$\u = \left(\E,m,h,j,\dots\right)^T\in\Real^{21}$ is the vector of all
dynamic variables of the model and $\hat{\v D}=\diag(D,0,0,\dots)$ is the
tensor of diffusion in which only the 
coefficient of the \sizedel{transmembrane} voltage $\E$ is nonzero. 
This simplified description focuses on the excitation and propagation
of \sizedel{electrical} impulses, while ignoring the 
effects due to geometry, 
anisotropy and heterogeneity of a real atrium. 

\begin{figure}[t]
\begin{center}
  \psfrag{lntau}[bl][bl][1][0]{$\ln(|\tau/\mathrm{ms}|)$}
  \psfrag{20}[Br][Br][1][0]{20}%
  \psfrag{15}[Br][Br][1][0]{15}%
  \psfrag{10}[Br][Br][1][0]{10}%
  \psfrag{5}[Br][Br][1][0]{5}%
  \psfrag{-5}[Br][Br][1][0]{-5}%
  \psfrag{0}[cc][cc][1][0]{0}%
  \psfrag{e-2}[Bl][Bl][1][0]{0.01}%
  \psfrag{e-1}[Bl][Bl][1][0]{\hspace*{0.2em}0.1}%
  \psfrag{e0}[Bc][Bc][1][0]{1}%
  \psfrag{e1}[Br][Br][1][0]{10}%
  \psfrag{e2}[Br][Br][1][0]{100}%
  \psfrag{tms}[Bc][Bc][1][0]{\hspace*{0.5em}$t$\,[ms]}%
  \psfrag{E}[tl][tl][1][0]{$\tau_E$}%
  \psfrag{fast}[tl][tl][1][0]{fast}%
  \psfrag{slow}[tl][tl][1][0]{slow}%
  \psfrag{aa2}{$10^{-3}$}   % it was 10^{-2} in your version - vnb
 \psfrag{aa1}{$10^{-2}$}  
\psfrag{a0}{$10^{0}$}  
\psfrag{a1}{$10^{1}$}  
\psfrag{a2}{$10^{2}$}
\psfrag{a}{(a)}
\psfrag{b}{(b)}
\psfrag{c}{(c)}
\psfrag{d}{(d)}
\psfrag{INa}{$\INa/10$}
\psfrag{Iout}{$I_{\mathrm{out}}$}
\psfrag{Iin}{$I_{\mathrm{in}}$}
\psfrag{I}{$I$}
\psfrag{mi}{$\mbar^3$}
\psfrag{hi}{$\hbar$}
\psfrag{V}{$\E$ [mV]}
\psfrag{t}{$t$ [ms]}
\psfrag{-2000}{\hspace{-2mm}-2000}
\psfrag{-1500}{\hspace{-2mm}-1500}
\psfrag{-1000}{\hspace{-2mm}-1000}
\psfrag{-500}{\hspace{-2mm}-500}
\psfrag{0}{\hspace{-2mm}0}
\psfrag{500}{\hspace{-2mm}500}
\psfrag{1000}{\hspace{-2mm}1000}
\psfrag{-100}{\hspace{-2mm}-100}
\psfrag{-50}{\hspace{-2mm}-50}
\psfrag{0}{\hspace{-2mm}0}
\psfrag{50}{\hspace{-2mm}50}
\psfrag{1.0}{\hspace{-1mm}1.0}
\psfrag{0.0}{\hspace{-1mm}0.0}
\psfrag{0.5}{\hspace{-1mm}0.5}
\psfrag{-80}{-80}
\psfrag{-60}{-60}
\psfrag{-40}{-40}
\psfrag{-20}{-20}
\hspace*{-5mm}
\begin{minipage}[]{6cm}
\epsfig{file=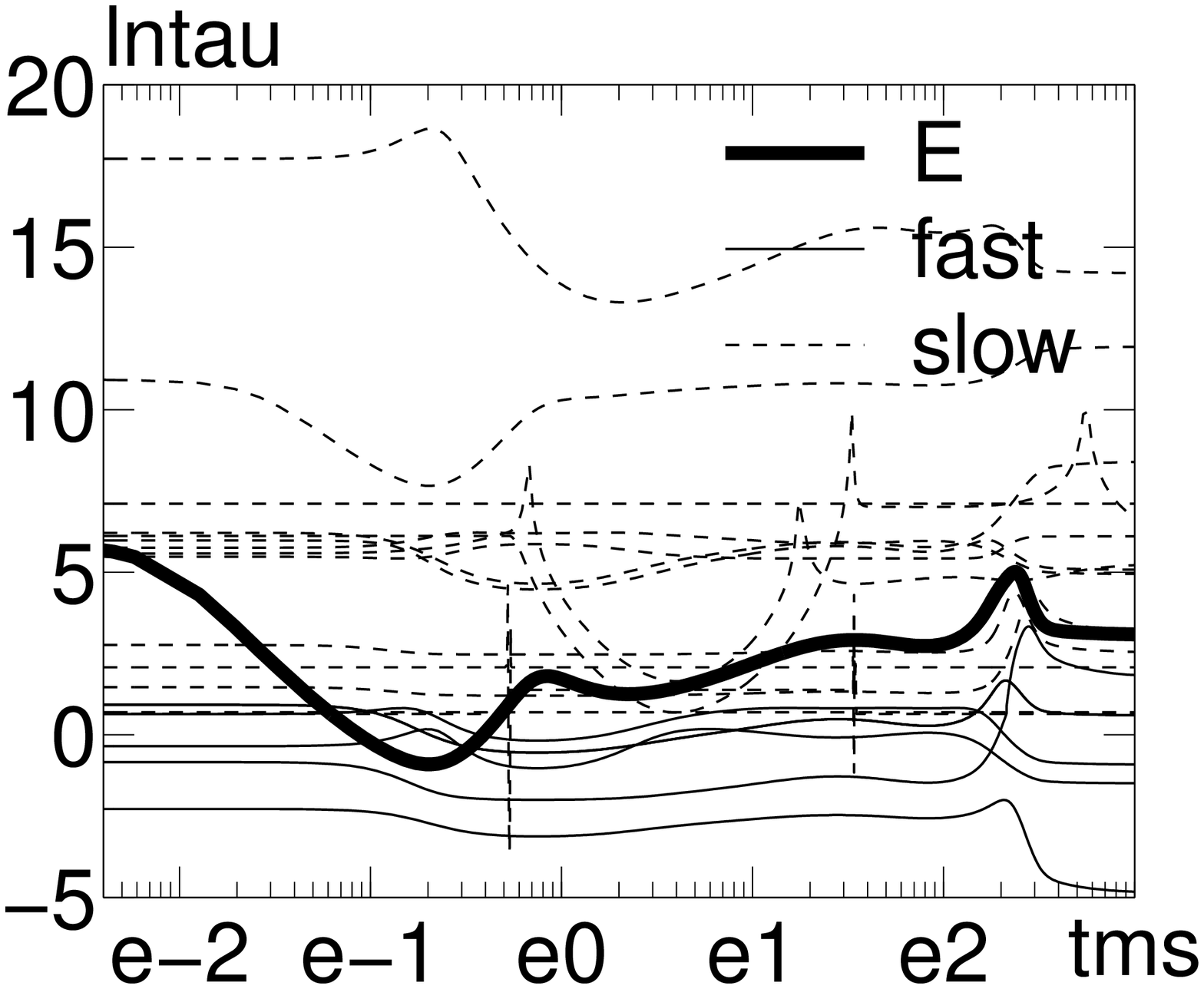,width=6cm,clip=} \\[-4mm]
\hspace*{0mm}\epsfig{file=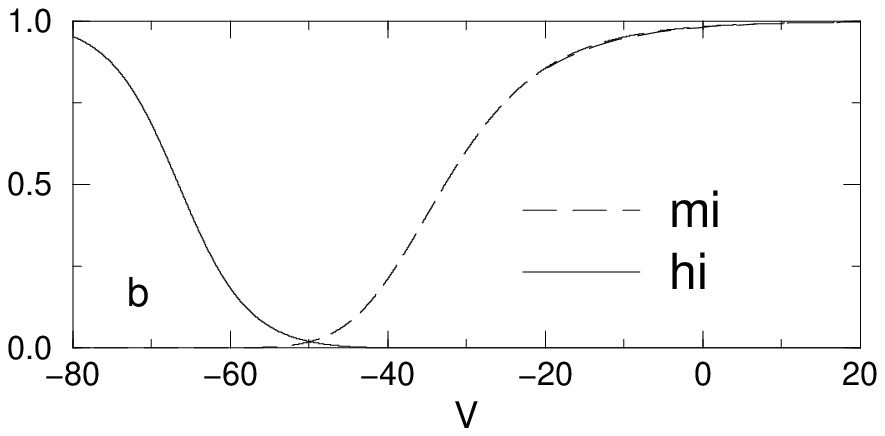,height=5.7cm,angle=-90,clip=} 
\end{minipage}
\begin{minipage}[]{6cm}
\vspace*{3mm}
%/export/home/volkenstein/simitev/WORK/Cour/V2.c VandCur.x
\epsfig{file=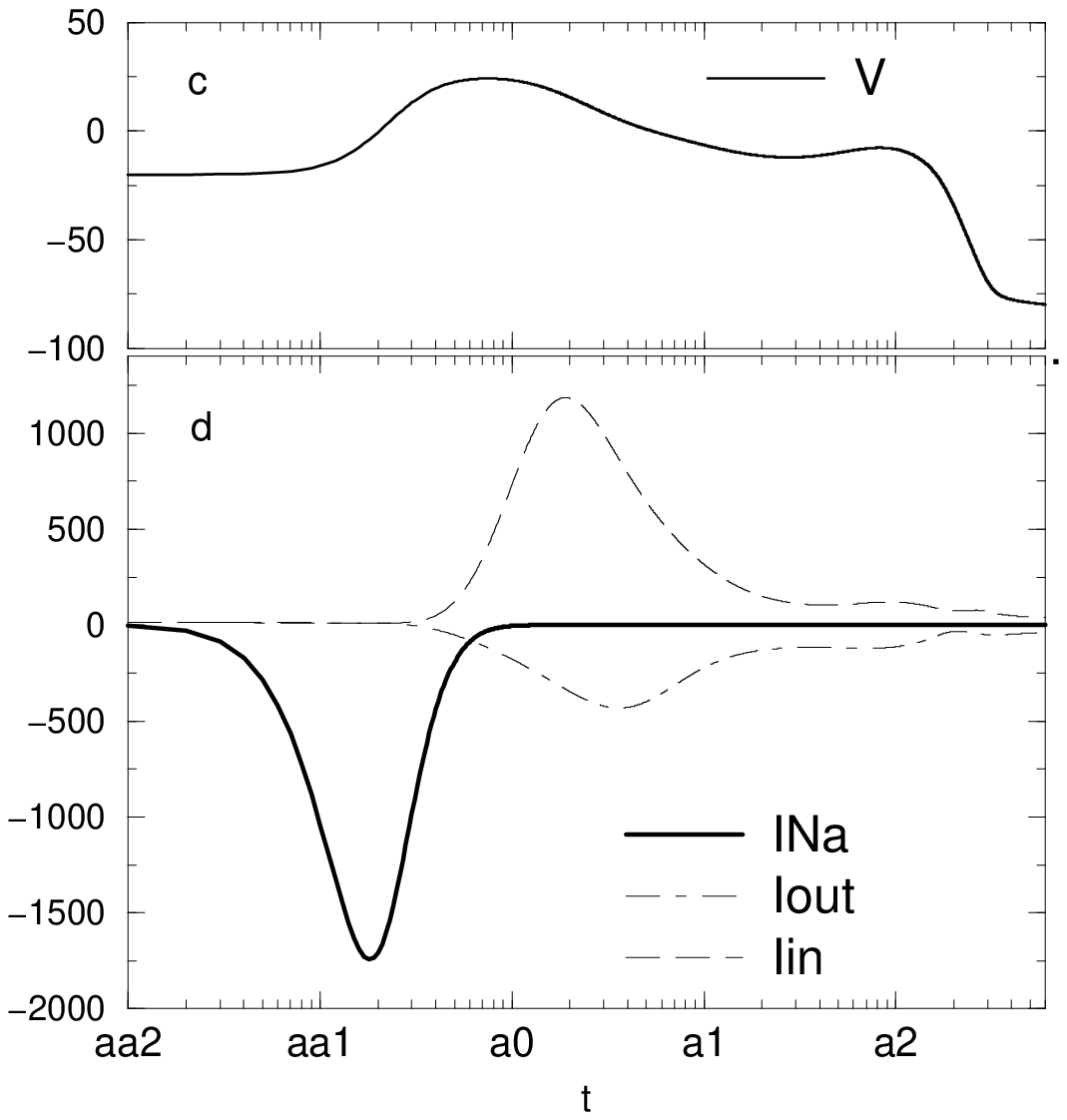,height=6cm,width=7.9cm,angle=-90,clip=} 
\end{minipage}
\end{center}
\caption[]{Asymptotic properties of the atrial \sizerepl{model of
  Courtemanche \etal{}}{CRN model} \cite{CRN98}. (a) Time scale functions of dynamical
  variables vs. time. (b) Quasistationary values of the gating
  variables $\mbar$ and $\hbar$. (c) Transmembrane voltage $\E$ as a
  function of time. (d) Main  ionic currents:
  $\INa$ is the fast sodium current (shown scaled by a factor 0.1),
  $I_{\mathrm{in}} = 
  I_{\mathrm{b,Na}}+I_{\mathrm{NaK}}+I_{\mathrm{Ca,L}}+I_{\mathrm{b,Ca}}+I_{\mathrm{NaCa}}$
  is the sum of all other inward currents and 
  $I_{\mathrm{out}} =
  I_{\mathrm{p,Ca}}+I_{\mathrm{K1}}+I_{\mathrm{to}}+I_{\mathrm{Kur}}+I_{\mathrm{Kr}}+I_{\mathrm{Ks}}+I_{\mathrm{b,K}}$
  is the sum of all outward currents;
  the individual currents are described in \cite{CRN98}.
  The results are obtained 
  for a space-clamped version of the model at values of the parameters
  as given in \cite{CRN98}.  A typical AP was triggered by initialising the
  transmembrane voltage to non-equilibrium value of $V=-20$~mV.
}
\label{f:0010}
\end{figure}

\subsub{Asymptotic reduction}
In order to reduce the dimension and complexity of the problem, we
perform  a formal analysis of the time scales  of dynamic
variables. For the system  \eqref{n:0010} we
define characteristic \textit{time scale} functions,
$
%\begin{gather}
\label{equ1}
\tau_i(u_1,\dots,u_{21}) \equiv \left|\left(\dd F_i/\dd u_i
  \right)^{-1}\right|
%\end{gather}
$
and compare their magnitudes obtained numerically for
a space-clamped version of the system as shown in figure \ref{f:0010}. 
The variables, whose time scales $\tau_i$ are relatively
small, are \textit{fast variables} since they change significantly during
the upstroke of a typical AP, while all other variables, whose time
scales $\tau_i$ are relatively large, are \textit{slow variables} because they
change only slightly during this period. Figure
\ref{f:0010}(a) demonstrates that the variables $\E$, $m$, $h$, $u_a$,
$w$, $o_a$, $d$ are fast variables comparable with the time scale of
the AP upstroke.

A specific feature of system \eqref{n:0010} is that of the various ionic
currents in the system only the sodium current $\INa$ is significantly
large during the AP upstroke, whereas other currents are small at this
stage as can be seen in figure \ref{f:0010}(d). Secondly, the fast
sodium current $\INa$ is only large 
during the AP upstroke, and almost vanishes otherwise, because either
gate $m$ or gate $h$ or both are nearly closed outside the upstroke 
since their quasistationary values $\mbar(\E)$ and $\hbar(\E)$ are small
there as illustrated in figure \ref{f:0010}(b).

To formalise the distinction between fast and slow terms we perform an
\emph{asymptotic embedding} \index{asymptotic embedding} of system \eqref{n:0010}. We introduce
an artificial parameter $\epsilon$ into the system so that 
for $\epsilon=1$ the original system is recovered, while in the limit
$\epsilon \rightarrow 0$ only the terms comparable with the time scale of
the AP upstroke are retained, 
\begin{align}
\hspace*{-6mm}
&\D_{T}{\E} = D \D_{X}^2 {\E} - \f{\left(\epsilon^{-1}\INa(\E,m,h,j) +
  \SIsmall(\E,\dots) \right)}{\CM},
                          \nonumber\\
&\D_{T}{m} = \frac{\big(\emb{\mbar}(\E;\epsilon)-m\big)}{\epsilon \, \tau_{m}(\E)}, \quad
         \mbar(\E;\epsilon) =\left\{
\begin{array}{l@{\extracolsep{3mm}}l} 
\mbar(\E), &  \epsilon =1, \\% \ne 0,\\
\Heav(\E-\Em), & \epsilon=0,\\
\end{array}\right. \nonumber\\
%\emb{\mbar}(\E;0)=M(\E)\Heav(\E-\Em),          \nonumber\\
&\D_{T}{h} = \frac{\big(\emb{\hbar}(\E;\epsilon)-h\big)}{\epsilon \, \tau_{h}(\E)}, \qquad
         \hbar(\E;\epsilon) =\left\{
\begin{array}{l@{\extracolsep{3mm}}l} 
\hbar(\E), &  \epsilon =1,\\ %  \ne 0,\\
\Heav(\Eh-\E), & \epsilon=0,\\
\end{array}\right. \nonumber\\
%         \emb{\hbar}(\E;0)=H(\E)\,\Heav(\Eh-\E),          \nonumber\\
&\D_{T}{y} = \frac{\big(\overline y(\E)-y\big)}{\epsilon \,
  \tau_{y}(\E)}, \nonumber \qquad \qquad y= \ua, w,\oa,d, \\
%\big
&\D_{T}{\U} = \v W(\E,\dots),                     \label{e:CRN}
\end{align}
where
$\Heav()$ is the Heaviside function, 
$\SIsmall()$ is the sum of all currents except the fast sodium current $\INa$,
the dynamic variables $\E$, $m$, $h$, $\ua$, $\oa$ and $d$
are defined in \cite{CRN98}, 
$\U=(j,o_i,\dots,\Nai,\Ki,\dots)^T$ is the vector of all other, slower variables, 
and $\v W$ is the vector of the corresponding right-hand sides. 
Novel features of the asymptotic embedding \eqref{e:CRN}, non-standard
in comparison with the theory of fast-slow systems 
\cite{Tikhonov-1952,Pontryagin-1957,Arnold}
are {\bf (a)} the introduction of the asymptotic 
factor $\epsilon^{-1}$ only at one term $\INa$ in the right-hand side
of the equation for $\E$ whereas the standard factor $\epsilon$ at the
derivative would be equivalent to factor $\epsilon^{-1}$ at the whole
right-hand side,
and {\bf (b)} that in the limit $\epsilon\to0$, functions
$\mbar(\E)$ and 
$\hbar(\E)$ have to be considered zero in  certain overlapping
intervals $\E\in(-\infty,\Em]$ and $\E\in[\Eh,+\infty)$, and
$\Eh\le\Em$, hence the representations
$\emb{\mbar}(\E;0)=\Heav(\E-\Em)$ and
$\emb{\hbar}(\E;0)=\Heav(\Eh-\E)$. 
These aspects, as applied to the fast sodium current, have been shown
to be crucial for the correct description of the propagation
block \cite{Biktashev-2002}. A more detailed discussion of the
parameterisation \eqref{e:CRN} can be found in
reference \cite{Biktasheva-etal-2005}.

The exact value of $D$ is not essential for the theoretical 
analysis, as its change is equivalent to rescaling of the
spatial coordinate. 
To operate with dimensional velocity, we
assume the values $D=0.03125$, 
as in \cite{Biktasheva-etal-2003,Biktasheva-etal-2005} and $\CM= 1
\mu$F cm$^{-1}$. 
We perform the scaling
${t}=\epsilon^{-1}T$, ${x}=(\epsilon D)^{-1/2}X$,
take the limit $\epsilon\to0$ and notice that the
equations for the variables denoted by $y$ in \eqref{e:CRN}
decouple from the voltage equation. Thus we arrive at the conclusion that
only the following three-variable system 
needs to be considered for a description of the propagation of an AP
front or its failure, 
\begin{subequations}
\label{e:0020}
\begin{align}
\label{e:0021}
& \dd_t \E = \dd_x^2 \E + \INabar(\E)\,j\,h\,m^3, \\
\label{e:0022}
& \dd_t h = \big(\Heav(\Eh-\E) - h\big)/\tau_h(\E), \\
\label{e:0023}
& \dd_t m = \big(\Heav(\E-\Em) - m\big)/\tau_m(\E).
\end{align}
\end{subequations}
%
%\sizedel
{In other words, we consider the fast time scale on which the upstroke
of the AP occurs, neglect the variations of slow variables
during this period as well as all transmembrane currents except $\INa$,
as they do not make significant contribution during this period and
replace $\mbar$ and $\hbar$ with zero when they are small.}
The parameters and functions in \eqref{e:0020} are defined as in
\cite{CRN98}, namely 
\begin{subequations}
\label{e:0030a}
\begin{align}
& \label{e:0030.1}
\INabar(\E) = g_{Na}(\ENa - \E), \\
&  \label{e:0030.2}
\tau_k(\E) = \big(\alpha_k(\E)+\beta_k(\E)\big)^{-1}, \qquad k=h, m,  \\
& \overline{k}(\E) = \alpha_k(\E)/\big(\alpha_k(\E)+\beta_k(\E)\big), \qquad k=h, m, \nonumber \\
& \alpha_h(\E) = 0.135\,e^{-(\E+80)/6.8}\,\Heav(-\E-40), \nonumber \\
& \beta_h(\E) = \left(3.56\,e^{0.079\E}+3.1\times 10^5\,e^{0.35\E}\right)\,
\Heav(-\E-40) \nonumber \\
&\hspace{2cm} + \Heav(\E+40)\,\big(0.13 (1+e^{-(\E+10.66)/11.1})\big)^{-1}, \nonumber \\
& \alpha_m(\E) = \f{0.32 (\E+47.13)}{1-e^{-0.1(\E+47.13)}}, \nonumber \\
& \beta_m(\E) = 0.08 e^{-\E/11}, \nonumber \\
& g_{Na}=7.8, \quad \ENa=67.53,  \quad \Eh=-66.66, \quad \Em=-32.7. \nonumber
\end{align}
\end{subequations}
Two new `gate threshold' parameters $\Eh$ and $\Em$ appear in the
system and are chosen from the conditions $\hbar(\Eh)=1/2$ and
$\mbar^3(\Em)=1/2$.   
As follows from the derivation, variable $j$, the slow inactivation gate
of the fast sodium current, acts as a parameter of the
model. It is the only one of all slow variables included in the vector $\U$ 
that affects our 
fast subsystem. We say that it describes the \emph{excitability} \index{excitability} of the
tissue. 

\subsub{Travelling waves and reduction to ODE}
We look for solutions in the form of a front propagating with a
constant speed and shape. So we use the ansatz $F(z) = F(x+ct)$ for $F=\E,h, m$ 
where 
$c$ is the 
dimensionless wave speed of the front, related to the dimensional
speed $C$ by  $c=(\epsilon/D)^{1/2}C$.
Then equations \eqref{e:0020} reduce to 
\begin{subequations}
\label{e:0030}
\begin{align}
\label{e:0031}
& \E'' = c \E' - \INabar(\E)\,j\,h\,m^3, \\
\label{e:0032}
& h' =  \big(c\,\tau_h(\E) \big)^{-1} \big(\Heav(\Eh-\E) - h \big), \\
\label{e:0033}
& m' =  \big(c\,\tau_m(\E) \big)^{-1}  \big(\Heav(\E-\Em) - m \big),
\end{align}
\end{subequations}
where the boundary conditions are given by 
\begin{subequations}
\label{e:0040}
\begin{align}
& \E(-\infty) = \Ealpha, & & h(-\infty)=1, & & m(-\infty)=0,    \label{e:0040a}\\
& \E(+\infty) = \Eomega,   & & h(+\infty)=0, & & m(+\infty)=1.  \label{e:0040b}
\end{align}
\end{subequations}
Here $\Ealpha$ and $\Eomega$ are the pre- and post-front voltages,
and $\Ealpha < \Eh < \Em < \Eomega$. 

Equations \eqref{e:0030} represent a system of fourth order so its
general solution  depends on four arbitrary constants. Together with
constants $\Ealpha$, $\Eomega$ and $c$ this makes seven constants to
be determined from the six boundary conditions in \eqref{e:0040}. 
Thus, we should have a one-parameter family of solutions, \ie{}
one of the parameters $(\Ealpha, \Eomega, c)$
can be chosen arbitrary from a certain range.

\begin{figure}[t]
%/export/simitev/WORK/IBVproblem/11num3var/FD/EXPLORE/CRN.vs.NUM.vs.ANL/plot05.x
\begin{center}
\psfrag{E}{\footnotesize $E$}
\psfrag{Z}{\footnotesize $Z$}
\psfrag{m}{\footnotesize $m$}
\psfrag{h}{\footnotesize $h$}
\psfrag{h,m}{\footnotesize $h,m$}
\epsfig{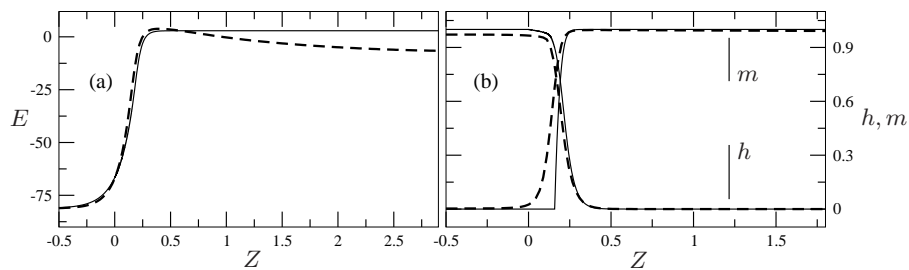}
\end{center}
\caption[]{
\reva[R6]{  (a) The AP potential and (b) the gating variables $h$ and $m$ as
  functions of the travelling wave coordinate $Z=z\sqrt{D}$. The
  solution of the \sizerepl{Courtemanche \etal{} model}{CRN model} \cite{CRN98} is given by
  broken lines and that of the three-variable model of \eqref{e:0030} by
  solid lines.  The prefront voltage and the excitation parameter in
  \eqref{e:0030} are chosen as $\Ealpha=-81.18$ mV and $j=0.956$,
  respectively and correspond to the equilibrium values in the
  realistic model. The gates $h$ and $m$  are indicated in the
  plot. The iterative analytical solution \eqref{e:0040:i0090} is
  indistinguishable from the  numerical solution of \eqref{e:0030}
  after 30 iterations.}}
\label{f:0020}
\end{figure}

\subsub{Comparison of the three-variable model \eqref{e:0030} with the
realistic \sizerepl{model of Courtemanche \etal{}}{CRN model} \cite{CRN98}}
\reva[R6]{
The simplified three-variable model \eqref{e:0020} and its ODE version
\eqref{e:0030} provide an excellent approximation to the fronts of the
action potential in human atrial tissue as demonstrated in figure
\ref{f:0020} where a comparison with the solution of the realistic 
\sizerepl{model of Courtemanche \etal{}}{CRN model} \cite{CRN98} is presented. As must be expected
the voltage in the simplified model remains constant after reaching
its post-front value $\Eomega$ while the voltage in the
\sizerepl{Courtemanche \etal{} model}{CRN model} assumes a shape typical for an action
potential. To quantify further the comparison between the two models
below we list
the values of the wave speed, the post-front voltage and the maximum
rate of AP rise. For the realistic model \vnba{\cite{CRN98}}
these values are $C=0.2824$
mm/ms, $\Eomega=3.60$ mV and $(\de{\E}/\de t)_{max}=173.83$ V/s. The
respective values for the simplified model \eqref{e:0030} are
$C=0.2372$ mm/ms, $\Eomega=2.89$ mV and $(\de{\E}/\de t)_{max}=193.66$
V/s. The relative errors made by the simplified model in estimating the
wave speed and the  the maximum rate of AP rise are 16\% and 11\%,
respectively, and  the absolute error in estimating the post-front
voltage is $-0.7$ mV.}  

\reva[R2]{
We recall that our main motivation for the derivation of the
three-variable simplified model was to reproduce the realistic front
dissipation behaviour of atrial tissue as it appears for example in the
\sizerepl{model of Courtemanche \etal{}}{CRN model} \cite{CRN98}. The
third column of figure \ref{f:0000} 
illustrates the dissipation of a propagating front in equations
\eqref{e:0020} in response to a temporary block of excitability. It is
observed that the dissipation behaviour of the simplified system 
resembles the one of the realistic \sizerepl{model of Courtemanche \etal{}}{CRN model}
In this aspect of the behaviour our model is superior to the simplified
models of FitzHugh-Nagumo type. }

\section{Iterative analytical solution}
\label{sec:3}
\subsub{Exact solution for $\mathbf{\E \leq \E_m}$}
For $\E \leq \E_m$ the reduced model \eqref{e:0030},\eqref{e:0040a} has a
two-parameter family of exact solutions,
  with parameters $\Ealpha$ and $c$.
Since the boundary condition $m(-\infty)=0$ is an equilibrium point of
\eqref{e:0033},  $m(z)=0$ remains a solution for all $z
\leq \xi$. It follows that \eqref{e:0031} is a constant-coefficient
linear homogeneous equation in this interval,
 and its solution 
\begin{equation}
\label{e:0040:i0010}
\E(z) = \Ealpha +(\E_h-\Ealpha)\, e^{c\,z},
\end{equation}
satisfies boundary conditions
$\E(-\infty)=\Ealpha$,
$\E(0)=\Eh$ and $\E(\xi)=\E_m$, provided that the internal boundary point $\xi$
is located at
\begin{equation}
\label{e:0040:i0030}
\xi = \f{1}{c}\ln\!\left(\f{\Em-\Ealpha}{\Eh-\Ealpha}\right).
\end{equation}
Finally, $h(\E)=1$ is a solution of \eqref{e:0032} with
boundary condition $h(\Ealpha)=1$ in the interval $\E \leq \E_h$,
because it belongs to its equilibrium set. In the interval $\E \geq \E_h$,
\eqref{e:0032} can be re-written in the form
\begin{equation}
\label{e:0040:i0040}
\f{\de}{\de \E}(\ln h)  = - \f{1}{c^2\,(\E-\Ealpha)\,\tau_h(\E)},
\end{equation}
and its solution can be immediately obtained, 
\begin{equation}
\label{e:0040:i0050}
h(\E) = \exp \left( - \f{1}{c^2} \int_{\E_h}^{\E} \f{\de \E}{(\E-\Ealpha)\,
    \tau_h(\E)} \right). 
\end{equation}

\subsub{Approximate solutions for $\mathbf{\E \geq \E_m}$}
For $\E \geq \E_m$, we rewrite \eqref{e:0030} as
\begin{subequations}
\label{e:0040:i0070}
\begin{align}
& \f{\de}{\de z} \left( \f{\de \E}{\de z}
  e^{-cz}\right) = f(z)\, e^{-cz},\\
& \f{\de}{\de z} \left( \ln h \right) = - \f{1}{c\, \tau_h(z)}, \\
& \f{\de}{\de z} \big( \ln(1-m) \big) = - \f{1}{c\, \tau_m(z)},
\end{align}
\end{subequations}
where
\begin{equation}
  f(z) = -\INabar\big(\E(z)\big)\,j\, h(z)\, m^3(z) .
\end{equation}
The boundary conditions are
\begin{subequations}
\label{e:0040:i0080}
\begin{align}
& \E(\xi) = \E_m,                       \label{e:0040:i0080:a}\\
& \E'(\xi)=c\,(\Em-\Ealpha),            \label{e:0040:i0080:b}\\
& h(\xi) = h_0 = \exp \left( - \f{1}{c^2} \int_{\E_h}^{\E_m} \f{\de \E}{(\E-\Ealpha)\, \tau_h(\E)} \right),  
                                        \label{e:0040:i0080:c}\\
& m(\xi)=0,                             \label{e:0040:i0080:d}\\
& \E'(\infty) = 0.                      \label{e:0040:i0080:e}
\end{align}
\end{subequations}
This problem is equivalent to the following system of integral equations
\begin{subequations}
\label{e:0040:i0090}
\begin{align}
&\hspace{0mm}
\label{e:0040:i0091}
 m(z) = 1 -\exp\left( - \f{1}{c} \int_{\xi}^{z} \f{\de \sigma}{\tau_m(\E(\sigma))}\right), \\ 
&\hspace{0mm}
\label{e:0040:i0092}
 h(z) = h_0 \exp \left( - \f{1}{c} \int_{\xi}^{z} 
  \f{\de \sigma}{\tau_h(\E(\sigma))}\right), \\
&\hspace{0mm}
\label{e:0040:i0093}
\E(z)  =\Em
    - \frac{1}{c} \left[\int\limits_{\xi}^{z} f(\sigma)
 \left(1-e^{c(\xi-\sigma)}\right) \d\sigma -  \left(e^{cz}-e^{c\xi}\right)
 \int\limits_{z}^{+\infty} f(\sigma)  e^{-c\sigma} \d\sigma \right] \\
&\hspace{0mm}
\label{e:0040:i0094}
   c = \frac{1}{\E_h-\Ealpha}   \int_{\xi}^{+\infty} f(\sigma)\,
   e^{-c\sigma}  \de \sigma .
\end{align}
\end{subequations}
The last equation \eq{e:0040:i0094} imposes an additional
relationship between parameters $\Ealpha$ and $c$,
so the ultimate solution depends only on one arbitrary
parameter, say $\Ealpha$, which in reality may be determined
by the pre-history of the medium through which the excitation front propagates.

The system \eq{e:0040} can be solved by iterations, starting from a suitable
initial approximation.
The iterations converge for a 
% wide range of reasonable 
\vnba{number of various}
initial approximations, see figure
\ref{f:0030}.
In section \ref{sec:4} we discuss two selected initial approximations.
The first of
them, {\bf A1},
is favourable from a numerical point of view. The
second one, {\bf A2}, is important in the context of our recent work
\cite{Simitev06b} since it leads to an even further formal
simplification of  problem \eqref{e:0030} to a system which allows
exact solution and extensive analytical study.

\section{Selected  initial approximations}
\label{sec:4}

\subsub{A1: The small-diffusion initial approximation}
A simple initial approximation may be obtained by considering a
space-clamped version of \eqref{e:0030}
corresponding to the limit  $D \rightarrow 0$ of very small
constant of diffusion in equations \eqref{e:CRN} 
(note that this is applied only to the solution in the interval $\E \geq \Em$)
and replacing the
gating variable $m$ with its quasi-stationary value $\mbar$ which in
our asymptotic limit 
% corresponds to the Heaviside function $\theta(\E-\Em)$. 
and for $\E\geq\Em$ equals 1.
Hence, an initial approximation may be chosen to
satisfy 
\begin{subequations}
\label{pb:0030d1}
\begin{align}
\label{pb:0031d1}
& \f{\d E}{\d t} =  \INabar(\E)\,j\,h, \\
\label{pb:0032d1}
&  \f{\d h}{\d t} =  -h / \big(\tau_h(\E) \big).
\end{align}
\end{subequations}
This system can be solved in quadratures, 
\begin{subequations}
\label{pb:0030d2}
\begin{align}
\label{pb:0031d2}
& h(\E)= h_0 - \int_{\E_0}^\E \f{\d \E}{\INabar(\E)\,j\,\tau_h(\E)}, \\
\label{pb:0032d2}
& t = \int_{\E_0}^\E \f{\d \E}{\INabar(\E)\,j\, h(\E)},
\end{align}
\end{subequations}
where the initial conditions
$\E_0$ and $h_0$ are % suitable initial conditions, for instance,
given by \Eq{e:0040:i0080:a} and \Eq{e:0040:i0080:c}.
% to ensure approximate matching with the exact solution in the interval $\E \leq \Em$.

\subsub{A2: The `caricature' initial approximation}
An even simpler initial approximation is 
\begin{equation}
\E=\E^{\{0\}}= \mbox{const.} 
\end{equation}
In this case functions of voltage ${\INabar(\E)}$, 
${\tau_h(\E)}$ and ${\tau_m(\E)}$ 
take constant values
${\INabar(\E^{\{0\}})}$, ${\tau_h(\E^{\{0\}})}$ and
${\tau_m(\E^{\{0\}})}$, respectively.
% Then equations 
Then quadratures 
\eqref{e:0040:i0090} %
% can be simplified 
for the results of the first iteration are
% and the next approximation can be 
obtained in explicit formulae;
moreover, \eqref{e:0040:i0094} is resolved explicitly.
These explicit formulae have been reported in our recent work 
\cite{Simitev06b} (see formulae (9) of that paper). 
There this piecewise-linear simplification was considered only as an
arbitrary ``caricature''
with the purpose of merely analysing qualitative features of the solution set. 
Here we note that it actually appears naturally  as a step the iterative procedure
leading to a numerically accurate solution.

\reva[R6]{The iterative analytical solution \eqref{e:0040:i0090} obtained from
these initial approximations, is indistinguishable from the
numerical solution of \eqref{e:0030} after some 30 iterations and has
the shape of a travelling front as shown in figure \ref{f:0020}.}  
\begin{figure}[t]
\begin{center}
%\hspace*{-5mm}
%/export/home/volkenstein/simitev/WORK/IBVproblem/11analit3var/NUMERICAL/temp/bwconv02.agr
% \epsfig{file=bwconv.eps,width=11cm,clip=}
%\mypsfrag{cc}{\hspace{0mm}0.2327}
%\mypsfrag{Vo}{\hspace{-1mm}2.89}
%\mypsfrag{1}{1}
%\mypsfrag{10}{10}
%\mypsfrag{100}{100}
%\mypsfrag{0}{0}
%\mypsfrag{10}{10}
%\mypsfrag{-10}{-10}
%\mypsfrag{-30}{-30}
%\mypsfrag{0.05}{0.05}
%\mypsfrag{0.15}{0.15}
%\mypsfrag{0.25}{0.25}
%\mypsfrag{0.35}{0.35}
%\mypsfrag{C}{\hspace{-5mm}$C$ [mm/ms]}
%\mypsfrag{V}{\hspace{-5mm}$\Eomega$ [mV]}
%\mypsfrag{n}{\hspace{-5mm}iterations}
\hspace*{-5mm}
\epsfig{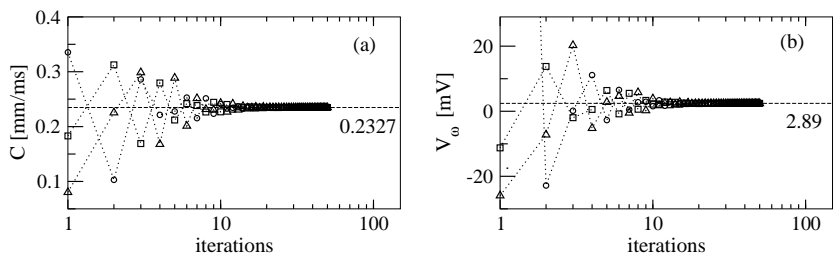}
\end{center}
\caption[]{ Convergence of the iterative solution
  \eqref{e:0040:i0090},
  starting from initial approximation \textbf{A1} (triangles),
       same as \textbf{A1} but with the equation for $m$ gate retained
  (squares) and   \textbf{A2} with $\E^{\{0\}}=\Em$ (circles).
%-  \reva[R5]{(c) and (d) Convergence of the iterative solution 
%-  \eqref{e:0040:i0090} starting from initial approximation
%-  \textbf{A2} with initial values $\E^{\{0\}}=-28, -30, -32, -34, -36,
%-  -38$ in order to demonstrate uniqueness. }
  The excitation parameter and the
  pre-front voltage are $j=0.9775$ and $\Ealpha=-81.18$~mV.
  The values on the right-hand side y-axis represent the 
  numerical  solution of the boundary value problem \eqref{e:0030}.
  }
\label{f:0030}
\end{figure}

\subsub{Convergence and uniqueness of the iterative solution
  \eqref{e:0040:i0090}}
\label{conv+uniq}
\reva[R5]{
\vnba{The iteration procedure produced by~\eqref{e:0040:i0090}
is nonlinear and non-monotonic,
and we do not have a rigorous proof of its convergence from 
any given initial approximation. Likewise, we do not have
a rigorous proof that the solution of the boundary-value
problem~\eqref{e:0040:i0080} is unique. 
However it is straightforward
to see that if the iterations converge, the result is a solution of the boundary
value problem, 
due to the above mentioned equivalence of \eqref{e:0040:i0080} and
the system of integral equations~\eqref{e:0040:i0090}.}}
Figure \ref{f:0030} illustrates that the first several iterations
obtained from \vnba{different initial approximations} 
% fluctuate quite randomly
oscillate
about the correct numerical solution of the problem % before
and
ultimately converge to it. % after about 25 steps. 
The ``small diffusion'' initial approximation {\bf A1} is 
% special 
particularly interesting
because 
% at the second step of the iterative procedure it leads to values which
% are relatively
the results of the second iteration are already very
close to the accurate numerical values for the wave speed
and the post-front voltage. Indeed, if the excitability parameter and
the pre-front voltage are chosen at their physiological resting values
$j=0.9775$ and $\Ealpha=-81.18$~mV, respectively \cite{CRN98}, at the
second step of the iterations the value of the dimensional wave speed is
$C=0.2255$~mm/ms which has only 20\% relative error compared to the
  value $0.2824$~mm/ms of the realistic ionic model \eqref{equ1} and 
5\% relative error compared to the 
value $0.2372$~mm/ms obtained numerically from equations
\eqref{e:0030}. Similarly the post-front voltage $\Eomega=-7.22$~mV
compares well with the value 3.36~mV of the \sizerepl{model of
Courtemanche \etal{}}{CRN model} \eqref{equ1}.
Clearly, the second iteration obtained from initial
approximation {\bf A1} introduces certain errors. Numerically,
however, it is immensely superior to any other numerical scheme
because it involves only a single evaluation of formulae \eqref{pb:0030d2} and a
two-fold evaluation of formulae \eqref{e:0040:i0090}. In addition, the dangers of
numerical divergence associated with many of the alternative numerical
schemes for solving problems \eqref{equ1} or \eqref{e:0030} are
avoided since the above expressions are mathematically well-behaved. 

\reva[R5]{
The two initial approximations discussed above are essentially
different. In the small-diffusion approximation the initial
guess for the voltage $V(z)$ is a function $V^0=v(z)$ while in the
`caricature' approximation it is a constant $V^0=$const. The
fact that two essentially different initial approximations give the
same limits is a strong indication that the iterative procedure leads
to a unique solution. To support this claim further 
\vnba{we have performed calculations starting from initial approximation
\textbf{A2} with initial values $\E^{\{0\}}=-28, -30, -32, -34, -36,
-38$. In all cases the iterations converged to the same solution,
in a similar manner to those shown on figure~\ref{f:0030}.
}
%-we present in
%-figure \ref{f:0030} the convergence to the same limit from several
%-other values of the `caricature' initial approximation. 
The 
small-diffusion initial approximation is not discussed here because
its initial guess $V^0=v(z)$ is the unique solution \eqref{pb:0030d2}
of equations \eqref{pb:0030d1} and thus it does not 
% allow variations.
\vnba{depend on any arbitrary parameters.}

%-We cannot exclude the possibility that ``bad'' initial guesses for the
%-voltage which are not based on reasonable physiological assumptions
%-may lead to divergence of the iterative scheme. However, there is
%-little reason to make initial guesses other than these which are
%-suggested in the paper. Indeed, it is difficult to find an initial
%-guess which is simpler than approximation {\bf A2} where the voltage
%-is assumed constant. 
\vnba{So, although we cannot exclude the possibility that 
the iteration procedure may not converge from some ``bad'' initial guess,
the examples considered provide an evidence that 
a reasonable initial approximation always gives converging iterations, 
and the solution is unique.
}
}

\section{Discussion}

\reva[R4]{
We have presented an analytical approach to the description of the
speed and the structure of an excitation front in a model of human
atrial tissue~\cite{CRN98}. We have identified small parameters in the
realistic model and used asymptotic arguments to obtain a
simplified three-variable model of the excitation front.}  
Although we have explicitly used certain quantitative features
of atrial tissue \cite{CRN98}, the main properties used are generic
for cardiac excitation models so the approach should be applicable,
possibly with suitable modifications, to other cardiac equations
models too. 
\reva[R4]{
Our model takes the form of a nonlinear eigenvalue problem with a
piece-wise right-hand side defined over three voltage intervals.  This
model is solved explicitly in the first two intervals, and in the last
interval we have suggested an analytical iterative procedure capable
of producing 
%- an accurate solution 
\vnba{a solution with a good accuracy}
already after the first iteration.
The iterative procedure can be started from reasonably chosen
simple initial approximations and converges to a unique solution which
differs only within few percent from the solution of the realistic
ionic atrial model.
}

An important feature of our approach is that it is capable of
correctly describing the excitation propagation at reduced
excitabilities, up to a complete block of propagation via ``front
dissipation'' mechanism, which is completely
\sizerepl{contrary to}{unachievable by} traditional
analytical approaches based on FitzHugh-Nagumo \sizerepl{asymptotic structure}{type} of
equations. This aspect has been analysed in our earlier publications~\cite{%
  Biktashev-2002,%
  Biktashev-2003,%
  Biktasheva-etal-2005,%
  Simitev06b%
}. \reva[R7]{In particular, our recent study \cite{Simitev06b} is
devoted to a discussion of 
% possible medical applications 
\vnba{one practical application}
of our approach. There we have used 
\vnba{a numerical solution of}
the simplified three-variable model
\eqref{e:0030}  to propose a simple criterion for break-up and
self-termination of spiralling waves and have confirmed our
predictions by numerical simulations of the realistic model of
Courtemanche \etal{}. However, the important question of finding
an analytical solution of our simplified model \eqref{e:0030} was now
solved in \vnba{this} paper.} 
 
The possibility of obtaining numerically reasonable analytical
approximations to front solutions in realistic cardiac equations,
demonstrated  in this paper, opens the way for analytical description
and, possibly, a better understanding, of more complicated regimes in
excitable media, such as wave break-ups and spiral waves. 

This study is supported by EPSRC grant~GR/S75314/01.

%\bibliographystyle{asm}
%\bibliography{asm}

\end{document}

%% file: abbr.tex
\newcommand{\CM}{C_M}
\renewcommand{\D}{\partial}

\renewcommand{\d}{\mathrm{d}}

\newcommand{\emb}[1]{{#1}}
\renewcommand{\E}{V} % It was V in CRN-1998 and E in most of our prev papers
\newcommand{\Eh}{\E_h}
\newcommand{\Em}{\E_m}
\newcommand{\ENa}{\E_{Na}}
\newcommand{\Ealpha}{\E_{\alpha}}
\newcommand{\Eomega}{\E_{\omega}}

\newcommand{\Heav}{\theta}
\renewcommand{\hbar}{\Qstat{h}}
\newcommand{\INa}{I_{\mathrm{Na}}}
\newcommand{\INabar}{\overline{\INa}}

\newcommand{\Ki}{Ki}
\newcommand{\mbar}{\Qstat{m}}
\newcommand{\Nai}{Nai}
\newcommand{\oa}{o_a}
\newcommand{\Qstat}[1]{\overline{#1}}
\newcommand{\SIsmall}{{\Sigma_{I}^{\prime}}}

\newcommand{\ua}{u_a}

\newcommand{\mx}[1]{\mathbf{#1}}
\newcommand{\U}{\mx{U}}

\newcommand{\eq}[1]{Eq.~\ref{#1}}

\newcommand{\Eq}[1]{Eq.~\ref{#1}}

\newcommand{\citet}[1]{\cite{#1}}

\newcommand\etal{\mbox{\textit{et al.}}}
\newcommand\etc{etc.}
\newcommand\eg{e.g.}
\newcommand\ie{i.e.}

\newcommand{\Real}{\mathbb{R}}            % set of real numbers
\renewcommand{\u}{\vec u}
\newcommand{\diag}{\mathrm{diag}}         % diagonal matrix